\documentclass[sigconf]{acmart}
\usepackage[plain]{fancyref}
\usepackage[draft=true]{minted} 

\usepackage{booktabs}
\usepackage{tabularx}
\usepackage{microtype}
\usepackage{multirow}
\usepackage{siunitx}
\usepackage{subcaption}
\usepackage{url}
\usepackage{graphicx} 
\makeatletter
\def\mdseries@tt{m}             
\makeatother

\setcopyright{none}
\title{Chunked Lists versus Extensible Arrays for Text Inversion}

\author{David Hawking}
\orcid{0000-0002-3704-5398}
\affiliation{\institution{School of Computing, Australian National University}\city{Canberra}\country{Australia}}
\email{david.hawking@acm.org}

\author{Bodo Billerbeck}
\orcid{0000-0002-9311-8504}
\affiliation{\institution{}\city{Melbourne}\country{Australia}}
\email{billerbeck@gmail.com}

\date{May 2023}

\begin{document}

\maketitle



\section{Introduction/Abstract?}
In our 2017 work on in-memory list-based text inversion \cite{HawBil-2017-ADCS-listBased} we compared memory use and indexing speed of a considerable number of variants of chunked linked lists.  In the present work we compare the best performing of those variants (FBB -- dynamic Fibonacci chunking) with the extensible SQ array technique (SQA) presented in \cite{MofMac-2023-ADCS-extensibleArrays}.

Chunked lists allow efficient appending and efficient scanning but, unlike arrays, do not allow random access to the items in the list.  In the text inversion task we study, random access is not required.    

In fact, there are strong similarities between chunked lists and SQ arrays.  The SQ array segments correspond very closely to list chunks, except that they do not contain a NEXT pointer.   Instead, pointers to SQ array segments are stored within a dope vector which must grow as the SQ array is extended.  For simplicity let's use a single word ``component'' to refer to both SQ array segments and list chunks.

In both chunked lists and SQ arrays, the first components added are small but component size increases with the number of items in the list.  In FBB the list of chunks comprises runs in which the size of the components is constant.  The size of the components increases from run to run according to a Fibonacci sequence and the length of the runs also increases according to a Fibonacci sequence.

SQA also involves runs of components of the same length but the formula controlling the length of the runs and the formula determining the size of the components within a run are cleverly chosen to allow efficient random access to an individual item.  

Considering a postings list represented by $n$ components, the memory space required by both methods is the space required by the items plus the unused items in the last partially filled component.  In addition, both methods require space for $n$ pointers.   For FBB HEAD and TAIL pointers are stored for each word in the vocabulary structure. For SQ arrays, a  pointer is stored there to reference the dope vector and in general there will be a number of unused pointers at the end of the dope vector.   The simplest method of growing a dope vector results in wasted space.  A new larger vector is allocated, the elements of the smaller one are copied over, and the smaller vector is discarded.  A portion of the discarded dope vectors could potentially be re-used but this would reduce locality of reference.  

\begin{table*}[ht]    
    \caption{Memory use and time comparisons of SQA and FBB methods.  (Means of 5 runs on quiet machine.)}
    \label{tab1}
    \begin{tabular}{llrrrrrrr}
    \hline
    &&&&&Build&Traversal&Total&Indexing\\
    &&Records&$|V|$&Postings&Time&Time&Memory&Rate\\
    Corpus&Method&(M)&(M)&(M)&(sec.)&(sec.)&(MB)&(M/sec.)\\
    \hline
    Synth10B&SQA&1378&1.00&10,000&1627&102&58,892&5.80\\
    Synth10B&FBB&ditto&ditto&ditto&1373&93.5&58,146&6.81\\
    \hline
    clueTitles&SQA&272&16.6&1971&175&31.4&11776&9.52\\
    clueTitles&FBB&ditto&ditto&ditto&162&25.7&11595&10.5\\
    \hline
    WIKT&SQA&11.1&2.27&32.8&4.29&1.29&212&5.86\\
    WIKT&FBB&ditto&ditto&ditto&3.99&1.23&205&6.26\\
    \hline
    \end{tabular}
\end{table*}

\section{Analytical comparison}
For a postings list of length $l$, we define the cost of a method as the number of memory units required in excess of the number required by a single array of length $l$ (the oracular method).  We assume that the space required by a pointer is equal to the space required by a posting, i.e 1 memory unit.  In this analysis, we compute the method costs for postings lists of all lengths up to 1 million.  We assume that no postings are stored in the vocabulary table and that postings are not compressed.   We compute the mean cost over all the lengths.

\begin{figure}
\includegraphics[width=0.47\textwidth]{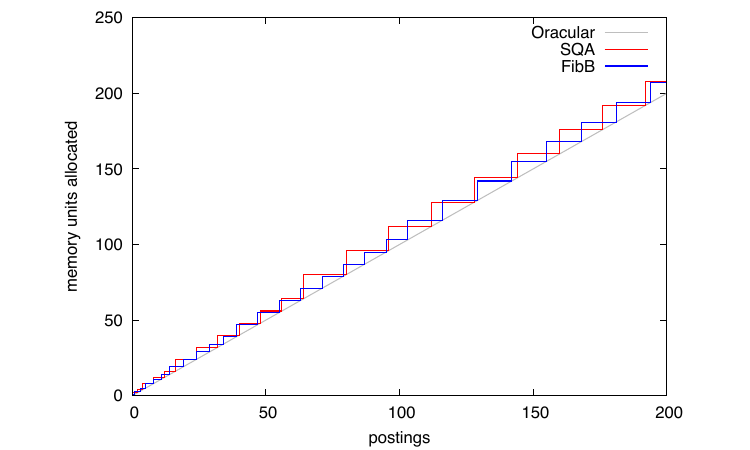}
\includegraphics[width=0.47\textwidth]{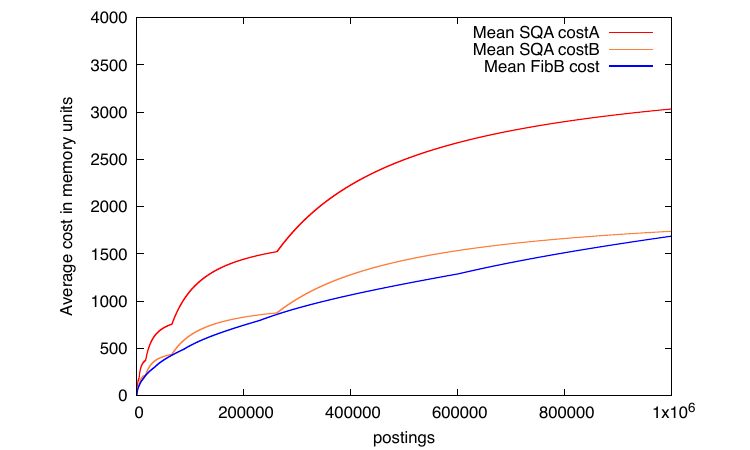}

\caption{L: The amount of space allocated for each method as the number of postings increases.  R: The average cost for each method as the number of postings increases.  Two costs are shown for SQA. A includes the cost of discarded dope vectors and B does not. Both A and B include the cost of 
internal fragmentation in the dope vector.}
\end{figure}

Figure 1 shows the allocations made by each method, and the average cost as the number of postings increases. After a million postings, FBB has allocated 2000 chunks with a maximum size of 1597, and incurred an average cost of 1688 memory units.  At the same point SQA has allocated 1488 chunks with a maximum size of 1024 and incurred an average cost of 3034 memory units, or 1739 memory units if it is (falsely) assumed that all discarded dope vectors can be re-used.

\section{Empirical comparison}
Experiments were performed on a 2023 MacBook Pro with Apple M2 Max chip, Ventura 13.4 OS, 96GB of "on-chip" RAM, and 2TB of SSD storage.   The retrieval system used was QBASHER,\footnote{Open-sourced at \url{https://bitbucket.org/davidhawking/qbasher}} coded in C11 and compiled with Apple clang version 14.0.3 and -O3 optimisation.

As before, our focus is on text corpora comprising large numbers of short texts, such as web page titles, quotations or song lyric lines.  We re-used the Wikipedia Titles (WIKT, 11 million records) from our previous work.  No longer having access to other corpora from that work, we extracted 272 million titles from various TREC sub-collections.  (clueTitles corpus.)  We also used SynthaCorpus\footnote{Open-sourced at \url{https://bitbucket.org/davidhawking/synthacorpus}} to generate a corpus (Synth10B) with 10 billion postings (1.37 billion records).

\section{Results, Discussion, and Conclusion}

Table \ref{tab1} shows memory use and time comparisons for SQA and FBB methods.  In QBASHER memory components and dope vectors is allocated sequentially within malloced units of 64 MB.  Total memory reported includes all malloced blocks.
On very large corpora for which QBASHER was designed, the SQA method uses approximately 1.3\% more memory than FBB.  This gap could be partially closed by re-using discarded dope vectors, at a small cost in time.  

Since the efficiency of coding of the two methods may differ, timing results should be regarded as preliminary.  The indexing rate for FBB was consistently faster: Synth10B by 17\%, clueTitles by 10\%, and WIKT by 7\%.

Curiously, indexing rates for WIKT and Synth10B are very similar but clueTitles indexes more than 60\% faster than WIKT for both FBB and SQA.   We plan to investigate this in future work.

In conclusion, the SQA method as implemented is observed to use slightly more memory and run a little slower than FBB but the differences would have negligible impact in practice.

\newpage

\begin{acks}
I thank Joel Mackenzie and Alistair Moffat for useful discussions and willingness to share code.
\end{acks}



\bibliographystyle{ACM-Reference-Format}
\bibliography{abbrevs-long,chunkedListsVsExtensibleArrays}

\end{document}